\documentclass[12pt,journal]{IEEEtran}
\usepackage{multirow}

\usepackage{xcolor}
\usepackage{lipsum}
\usepackage{subcaption}
\usepackage{graphicx}
\usepackage{tabularx,booktabs}
\ifCLASSINFOpdf
\else
\fi
\newif\ifcolormarker
\colormarkertrue

\begin{document}
%
\title{Network Digital Twin: Context, Enabling Technologies and Opportunities}

%
%
%

\author{\IEEEauthorblockN{Paul~Almasan, Miquel~Ferriol-Galmés, Jordi~Paillisse, José~Suárez-Varela, Diego~Perino, Diego~López, Antonio~Agustin~Pastor~Perales, 
Paul~Harvey, Laurent~Ciavaglia, Leon~Wong, Vishnu Ram, Shihan~Xiao, Xiang~Shi, Xiangle~Cheng, Albert~Cabellos-Aparicio, Pere~Barlet-Ros}

\vspace{0.3cm}
\small{\textbf{NOTE:} This work has been accepted for publication in the IEEE Communications Magazine.}\\
\vspace{0.05cm}
\scriptsize{© 2022 IEEE. Personal use of this material is permitted.  Permission from IEEE must be obtained for all other uses, in any current or future media, including reprinting/republishing this material for advertising or promotional purposes, creating new collective works, for resale or redistribution to servers or lists, or reuse of any copyrighted component of this work in other works.}
\vspace{-1cm}
\thanks{\textit{Paul~Almasan, Miquel~Ferriol-Galmés, Jordi~Paillisse, José~Suárez-Varela, Albert~Cabellos-Aparicio and Pere~Barlet-Ros are with Barcelona Neural Networking Center, Universitat Polit\`ecnica de Catalunya.}}
\thanks{\textit{Diego~Perino, Diego~López and Antonio~Agustin~Pastor~Perales are with Telefónica Research.}}
\thanks{\textit{Paul~Harvey, Laurent~Ciavaglia and Leon~Wong are with Rakuten Mobile.}}
\thanks{\textit{Vishnu Ram is an independent researcher.}}
\thanks{\textit{Shihan~Xiao, Xiang~Shi and Xiangle~Cheng are with Huawei Technologies Co., Ltd.}}
}

\maketitle

\begin{abstract}
The proliferation of emergent network applications (e.g., telesurgery, metaverse) is increasing the difficulty of managing modern communication networks. These applications entail stringent network requirements (e.g., ultra-low deterministic latency), which hinders network operators to manage their resources efficiently. In this article, we introduce the network digital twin (NDT), a renovated concept of classical network modeling tools whose goal is to build accurate data-driven network models that can operate in real-time. 
We describe the general architecture of the NDT and argue that modern machine learning (ML) technologies enable building some of its core components. Then, we present a case study that leverages a ML-based NDT for network performance evaluation and apply it to routing optimization in a QoS-aware use case. Lastly, we describe some key open challenges and research opportunities yet to be explored to achieve effective deployment of NDTs in real-world networks.
\end{abstract}


%
\IEEEpeerreviewmaketitle

\section{Introduction}

In the last years, the digital transformation of both society and industry has led to the emergence of novel network applications. These applications have complex requirements that cannot be easily met by traditional network management solutions at a reasonable cost, such as network over-provisioning or admission control. For example, novel forms of communication (e.g., AR/VR, holographic telepresence) require ultra-low deterministic latency, while recent industrial developments (e.g., Vehicular Networks) need to adapt to ever-changing network topologies in real-time. At the same time, the number of connected devices is growing massively, making modern networks highly dynamic and heterogeneous. As a result, communication networks are becoming increasingly complex and costly to manage.

Other industry sectors have recently adopted the digital twin (DT) paradigm \cite{8477101} to model complex dynamic systems. A DT can be understood as a virtual model of a physical object, system, or phenomenon that is represented in the digital world. The main advantage of DTs is that they can accurately model complex systems. Nowadays, DT applications include enabling smart manufacturing in Industry 4.0, improving the performance of complex engineering products (e.g., engine design) or modeling physical interactions (e.g., gravitational systems).

This article makes the case for the network digital twin (NDT) as a key enabler for efficient control and management of modern communication networks. NDTs can be applied to many fundamental networking applications. As an example, they allow network operators to perform online network optimization, what-if analysis, troubleshooting, or plan network upgrades considering the expected natural growth of the network. The interaction with the NDT does not require access to the real network, so the aforementioned operations can be performed without jeopardizing the physical network.

Recent machine learning (ML) models have shown outstanding capabilities for modeling complex systems. For example, in communication networks ML has already been successfully applied to network modeling~\cite{ferriol2022routenet}, 
traffic optimization in data centers~\cite{10.1145/3230543.3230551,10.1145/3452296.3472927}, network slicing~\cite{wang2020graph}, or resource allocation in wireless networks~\cite{eisen2020optimal}. In this context, we argue that modern ML techniques are a key enabler to build core components of the NDT. 

NDTs aim to achieve accurate data-driven network models operating in real-time~\cite{zhou-nmrg-digitaltwin-network-concepts-07, itu-standard}. In this vein, the use of ML enables training network models directly with real network data, avoiding the strong assumptions of analytical models (e.g., queueing theory). ML models can thus help achieve similar accuracy to traditional computationally-expensive modeling tools (e.g., packet-level simulation) while keeping a limited execution cost similar to lightweight analytical models. This allows network operators to accurately control the network at much shorter timescales.

There is a growing interest in the networking community in building NDTs. In particular, standards development organizations (SDO), such as the IETF or the ITU, have started to work on the definition of a NDT~\cite{zhou-nmrg-digitaltwin-network-concepts-07, itu-standard}. While their work focuses on defining the main concepts and interfaces of a NDT, this article focuses on the technologies and research challenges involved in implementing a ML-based NDT, complementing the work of SDOs. 

\section{The network digital twin}
\label{sec:archict}

\begin{figure}[!t]
  \centering
  \includegraphics[width=0.99\linewidth]{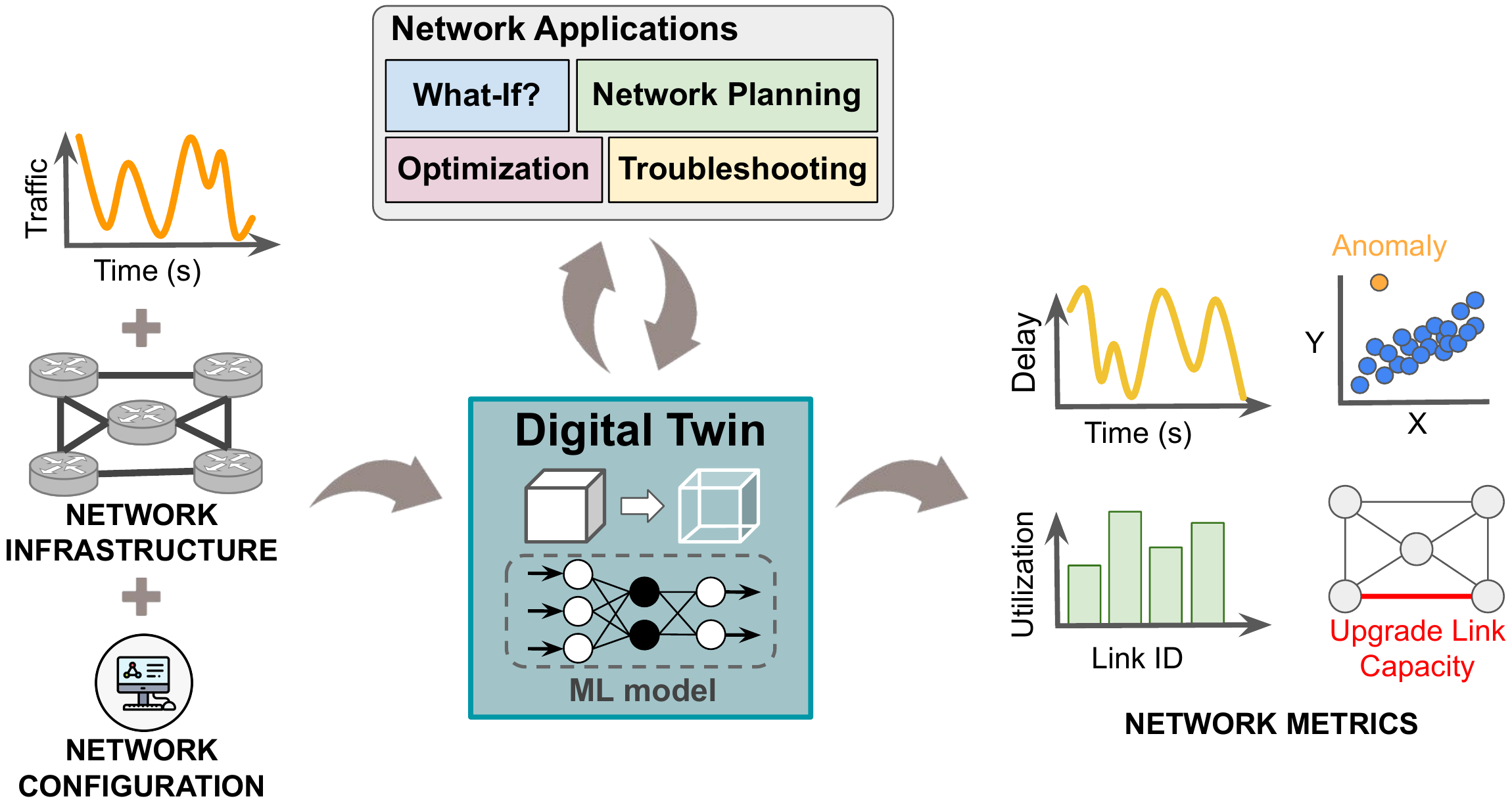}
  \caption{General network digital twin architecture.}
  \label{fig:DTNarchitecture}
\vspace{-0.2cm}
\end{figure}

NDTs are referred to as a new generation of network modeling tools that leverage ML techniques to build an accurate data-driven digital network representation~\cite{zhou-nmrg-digitaltwin-network-concepts-07, itu-standard}. To train these network models, we can use data from real-world networks, dedicated network testbeds, or network simulation tools. This data should be diverse enough to cover a wide representation of potential scenarios that the network operator wants to mimic (e.g., various congestion levels, link failures). In this context, recent deep learning (DL) techniques are of interest as they enable building accurate digital models of complex network environments~\cite{ferriol2022routenet,10.1145/3230543.3230551}.

Figure~\ref{fig:DTNarchitecture} presents the reference architecture of the NDT. The central component of the architecture is the DT, which implements a network model that mimics the physical network. This model takes as input a network state description (e.g., traffic, topology, routing, scheduling policies) and outputs some network-related metrics or features (e.g., utilization, delay, anomalies).

Since the NDT is a faithful copy of the real-world network, the network operator can test any input values, even if these values might cause service disruptions. This is because the NDT is executed in a safe environment isolated from the real-world network. The outputs can be of multiple types depending on the applications of the NDT (e.g., time series, link-level predictions, global network-level metrics). Note that the example depicted in Fig.~\ref{fig:DTNarchitecture} illustrates the case of a NDT applied to a fixed network, while analogous architectures could be applied to other kinds of networks, such as wireless/cellular networks. As an example, Table~\ref{table:dtn_applications} shows a description of some generic networking use cases that can take advantage of NDTs for efficient network control and management.

\begin{table}[!t]
\resizebox{\linewidth}{!}{
    \centering
    \begin{tabular}{m{1.5cm}m{6.0cm}}
    \hline
        \textbf{Application} & \textbf{Example Use case} \\ \hline \hline
        Trouble-shooting & Network operators can replicate past network scenarios with the NDT to find the root cause that produced a service disruption. \\ \hline
        What-If Analysis & The NDT acts as a \textit{safe sandbox} where different configurations can be applied to understand their impact on the network performance. \\ \hline
        Network Planning & The NDT can help estimate when an existing network will need a network upgrade. \\ \hline
        Anomaly Detection & When the behavior of the real-world network deviates from normal operational scenarios, the NDT can identify specific anomalies and point to some potential root causes. \\ \hline
    \end{tabular}
}
\caption{Overview of networking use cases enabled by NDTs.}\label{table:dtn_applications}
\vspace{-0.3cm}
\end{table}

\subsection{Leveraging machine learning to Build NDTs}

In this article, we argue that ML techniques are a key enabler to build core components of the NDT. Especially, recent DL models offer several advantages with respect to traditional network modeling tools (e.g., simulators, queueing theory). As an example, DL-based models have shown state-of-the-art performance when modeling fixed networks~\cite{ferriol2022routenet}, outperforming well-known analytical models based on queuing theory. In addition, they are easy to parallelize and have a low execution cost compared to traditional network simulation tools (e.g., OMNet++, \mbox{ns-3}).

Graph neural networks (GNN) are a DL-based architecture recently proposed by the ML community to model relational information~\cite{4700287}. GNNs capture graph dependencies using a message passing algorithm between the graph's entities (nodes and edges). Since communication networks are fundamentally represented as graphs, GNNs offer unique advantages for network modeling when compared to traditional NN architectures (e.g., multilayer perceptron, recurrent NN). In the last years, GNNs have demonstrated outstanding performance to solve a wide variety of network-related problems \cite{ferriol2022routenet,eisen2020optimal,10.1145/3452296.3472902, wang2020graph, 9651930}. In this context, GNNs may be a central technology to enable the construction of ML-based network models that can generalize to different network topologies, configurations, and traffic distributions.

\subsection{Network Optimization with the NDT}

\begin{figure}[!t]
  \centering
  \includegraphics[width=0.96\linewidth]{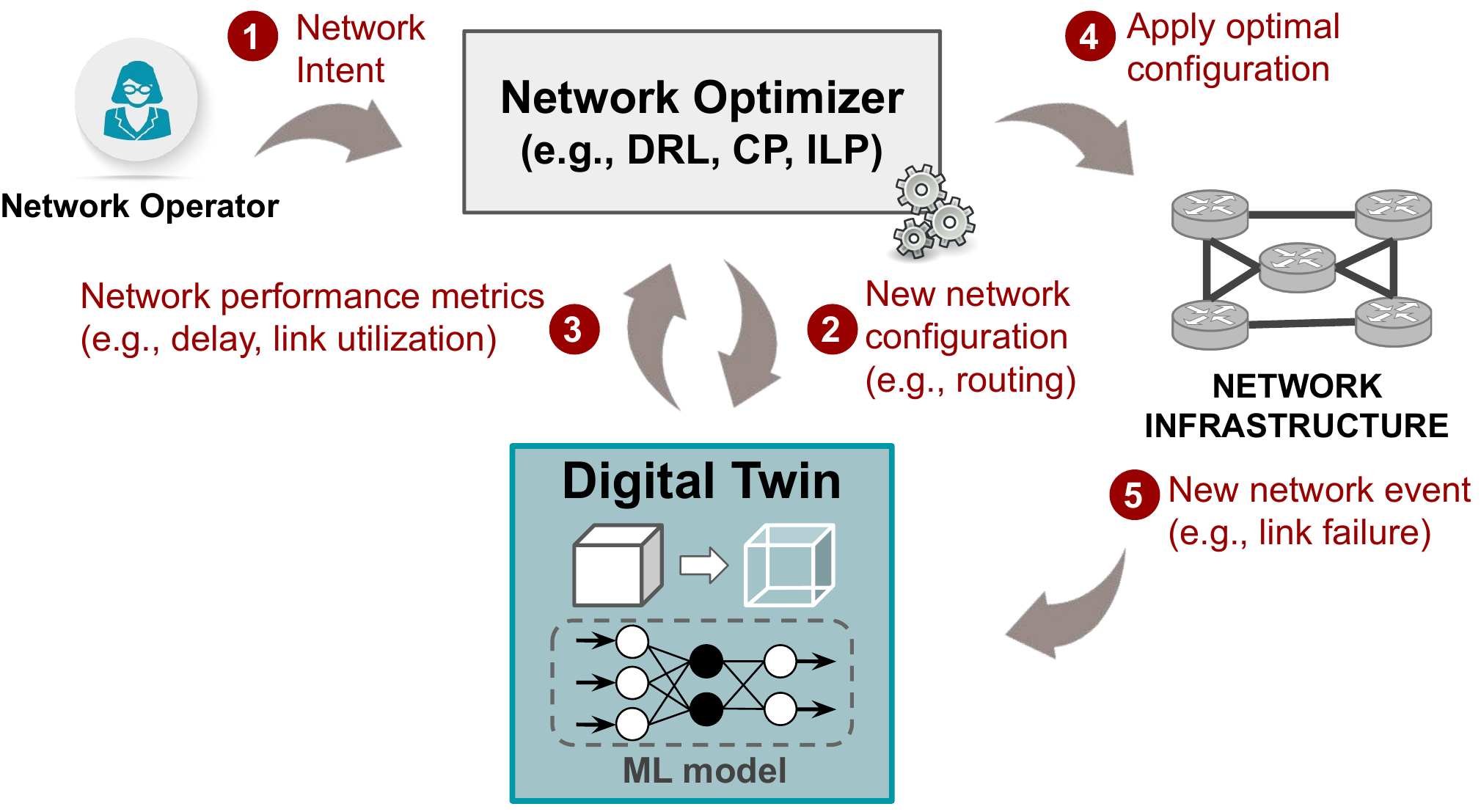}
  \caption{Network optimization with the NDT.}
  \label{fig:optimProcess}
\end{figure}

The NDT can be combined with a \textit{network optimizer} to solve different tasks (e.g., traffic engineering, network anomaly detection, network planning). Specifically, optimizers can use the NDT to obtain immediate network performance estimations during an optimization process. Figure~\ref{fig:optimProcess} summarizes this process. First, the network operator uses a declarative language to define the network requirements (e.g., load balancing). The optimizer is in charge of searching for the best network configuration that fulfills the predefined requirements (step~2). If the performance metrics from the NDT indicate that the solution is not good enough (step~3), then the network optimizer continues the search until a stopping condition is met. Lastly, the best solution found so far can be applied to the real network (step 4). Notice that the optimization process can be implemented as a closed-loop, with no human intervention required.

Real-world networks are highly dynamic as their traffic, applications, resource utilization and topology constantly changes over time. For example, physical links may break due to external factors, or network users can have different behavior patterns that cause difficult-to-predict spikes in the utilization of network resources. Therefore, to enable efficient network management, it is important for the optimizer to adapt to such changes in real-time. 

In this context, deep reinforcement learning (DRL) is a key technology that has shown great capabilities for efficient network operation in dynamic scenarios~\cite{10.1145/3452296.3472927, 10.1145/3230543.3230551, 9651930}. However, in complex optimization problems DRL often produces sub-optimal solutions. For example, in resource allocation problems it can be challenging to find the optimal network configuration that optimizes some performance metrics. This is because the solution space (i.e., the number of possible actions) might be very large, and more comprehensive exploration strategies are needed to find the optimal solution. Several works started combining DRL with traditional optimization methods (e.g., integer linear programming) to improve the optimization performance~\cite{10.1145/3452296.3472902}.

\section{Training the digital twin} 

Building a NDT requires collecting a dataset that contains relevant information of the network. The NDT's accuracy highly depends on the quality of the data, requiring the training dataset to contain a \emph{representative set} of samples with different network characteristics. For example, if the goal is to model the delay of network traffic, then the dataset has to include a wide range of network scenarios and its impact on the delay. This may include different routing configurations, topologies, scheduling and traffic loads. Likewise, the dataset should cover edge cases that may negatively impact the delay, such as link and interface failures, misconfigurations, highly congested scenarios, etc. 

Another important aspect to consider is, \emph{how do we generate this dataset?} Fundamentally, the dataset can be obtained from real-world networks, non-production dedicated testbeds, or simulation tools. However, generating such training sets in production environments may be impractical. As mentioned previously, the dataset must contain edge cases that may be unacceptable to reproduce in real-world networks as they could cause service interruptions. As a result, we envision that it is more practical to produce the training dataset in non-production 
environments, such as dedicated network testbeds or simulators. In these controlled environments, the network can be configured with different traffic profiles, failures, misconfigurations and errors, as well as covering a wide range of valid configurations without disrupting the normal operation of the network. 

The main challenge of generating the dataset is that the NDT has been trained in a specific network environment, but when deployed it has to operate on an unseen customer network. In other words, the NDT has to operate in scenarios that are not explicitly included in the training set. As an example, the topology and traffic profile of the customer network might be different from the ones seen during training in the controlled network environment. In the ML domain, the capability of a model to operate in unseen scenarios is referred to as \emph{generalization}.

\begin{table}[t]
  \centering
    \begin{tabular}{m{1.8cm}m{6.0cm}}
    \hline
        \textbf{Technology} & \textbf{Description} 
        \\ \hline \hline
        Recurrent NN & For each path, the RNN iterates over the sequence of links it traverses (represented by feature vectors). Link vectors are initialized with their capacity and traffic load. Then, a multilayer perceptron is used to compute the final delay per path. 
        \\ \hline
        Graph NN (RouteNet-E) & This GNN model represents the network as a set of paths and links. Then, it performs a message passing algorithm between the state of paths and links (represented by vectors), according to the input network topology and routing configuration~\cite{ferriol2022routenet}. A multilayer perceptron at the end predicts the final path delays.
        \\ \hline
        Queueing theory & Each link is modeled as a finite $M/M/1/b$ model. An iterative algorithm is repeated until the algorithm converges to a fixed point. Finally, the delay is computed for each path using standard queuing theory~\cite{kelly2011reversibility}. 
        \\ \hline
        Network simulator & Packet-level network simulator (OMNet++). It takes as input a network topology, a traffic matrix and a routing configuration, and simulates the mean per-packet delay for all paths. It is used as a ground truth for the experiments. 
        \\ \hline
    \end{tabular}
  \caption{Description of the baseline methods.}
  \label{tab:baslines}
 \vspace{-0.5cm}
\end{table}

\section{Case study: Performance Evaluation in Fixed Networks}
\label{sec:casestudy}

ML has already been validated for network modeling and optimization in many different scenarios (e.g., fixed networks~\cite{ferriol2022routenet}, data centers~\cite{10.1145/3230543.3230551,10.1145/3452296.3472927}, wireless networks~\cite{wang2020graph,eisen2020optimal}). In this section we present a case study that aims to analyze in more detail the application of a state-of-the-art ML-based NDT for performance evaluation in fixed IP networks. In addition, we perform some experiments where we leverage a ML-based NDT for routing optimization in a QoS-aware optimization use case.

\subsection{Predicting End-to-End Delay}

We take as a reference RouteNet-E~\cite{ferriol2022routenet}, a state-of-the-art GNN-based model that accurately predicts delays in networks. This model takes as input a network state description defined by: a network topology, a traffic matrix, and a routing policy. As a result, it mimics the network behavior and produces end-to-end delay predictions for all paths.

To train this model, we generate a dataset with 100,000 samples in topologies with 25-50 nodes simulated with an accurate packet-level network simulator (OMNet++). Then, we generate a test dataset with 500 samples from considerably larger topologies, with 50-300 nodes uniformly distributed. Network topologies are synthetically generated using the power-law out-degree algorithm, where the $\alpha$ and $\beta$ parameters have been extrapolated from real-world topologies of the Internet Topology Zoo repository~\cite{knight2011internet}. Traffic loads and link capacities are scaled to cover a broad range of congestion levels, with a maximum packet loss of $\approx$3\%. As reference baselines, we use a state-of-the-art analytical model based on queueing theory (QT) and a recurrent neural network (RNN). A detailed description of these baselines can be found in Table~\ref{tab:baslines}.

\begin{figure}[!t]
  \centering
  \includegraphics[width=0.90\linewidth, height=5cm]{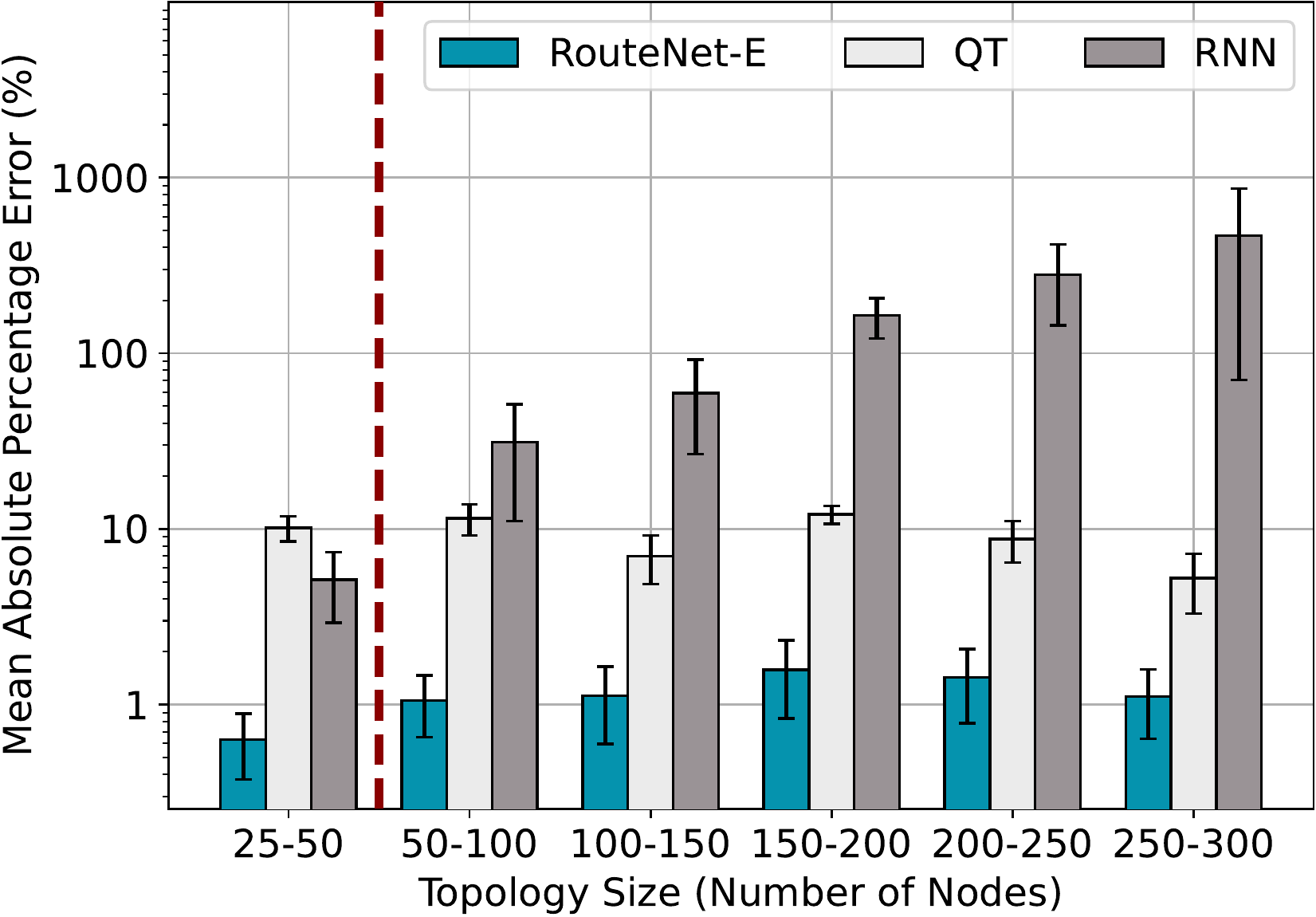}
  \caption{Comparison of several state-of-the-art methods for network performance evaluation.}
  \label{fig:barplot}
  \vspace{-0.5cm}
\end{figure}

Figure~\ref{fig:barplot} shows the evaluation results of the aforementioned models on topologies with up to 300 nodes. The y-axis represents the mean absolute percentage error of the predictions made by the different methods with respect to the ground truth labels produced by the network simulator. Error bars represent the 15/85 percentiles. If we look at the results in topologies of similar size to those of the training (25-50 nodes), we can observe that the two ML-based methods (RouteNet-E and RNN) achieve lower error than the analytical QT baseline, particularly in the case of RouteNet-E.

In this context, a potential limitation of ML-based solutions is that their accuracy is expected to drop when evaluated on out-of-distribution data. In this case, out-of-distribution data refers to topologies, traffic matrices and routing configurations different to those seen by the ML model during training. Figure~\ref{fig:barplot} shows the evolution of the prediction errors as we increment the network size with respect to the networks seen during training (with 25-50 nodes). We can observe that the RNN model significantly degrades its performance as networks become larger. In contrast, RouteNet-E shows a robust behavior when facing samples of considerably larger networks. This is thanks to its internal GNN-based architecture, which enables it to effectively model the relational information within networks and generalize well to larger topologies.

In addition, we compute the inference cost of all methods on off-the-shelf hardware (processor AMD Ryzen 9 3950X with 3.5GHz) on topologies with 250-300 nodes. As a result, we observe an average execution time of $\approx$0.16, $\approx$5.1 and $\approx$6.47 seconds for RNN, QT and RouteNet-E respectively, while the packet-level network simulator takes $\approx$3 hours and 39 minutes on average. 

Overall, the previous results show the potential benefits of modern ML models to produce performance estimates with similar accuracy to simulation methods, while keeping the limited cost of analytical models (e.g., QT), thus enabling fast operation.

\subsection{QoS-aware Routing Optimization}

\begin{figure}[!t]
  \centering
  \includegraphics[width=0.99\linewidth, height=3.5cm]{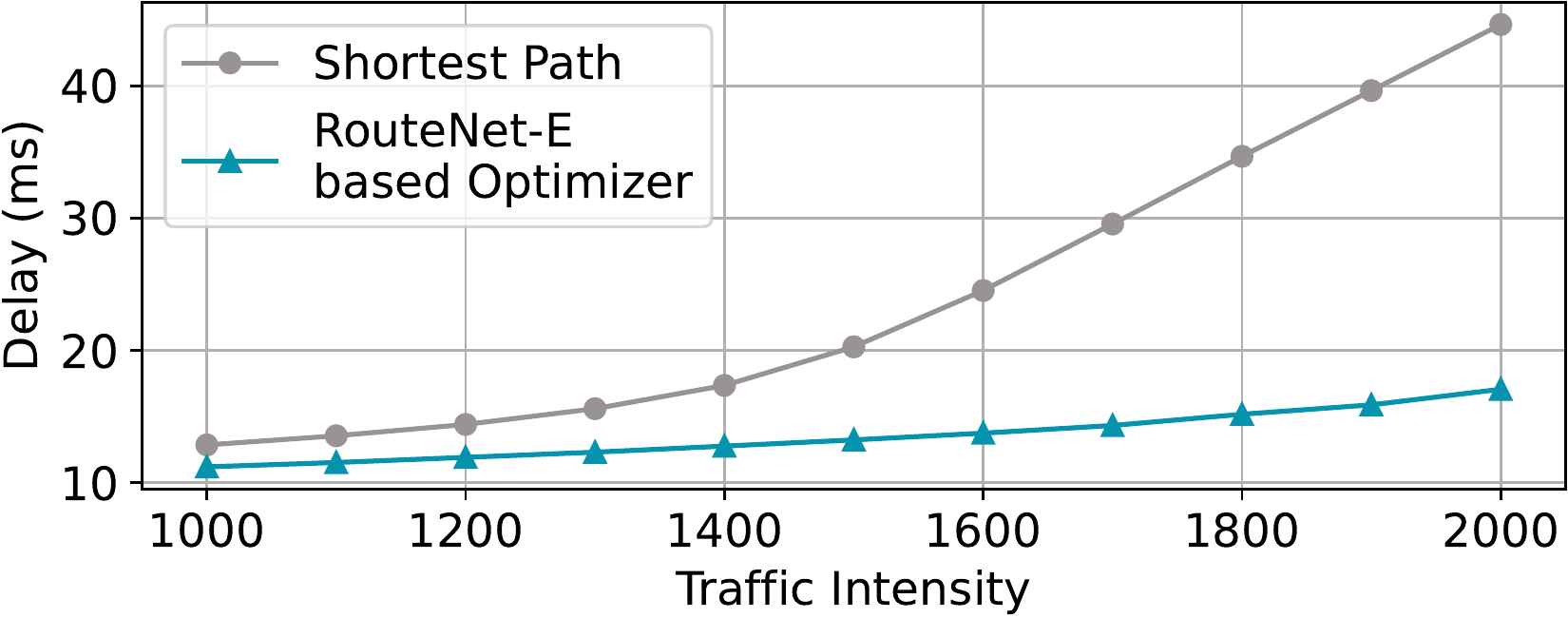}
  \caption{Delay-aware routing optimization using a ML-based NDT (RouteNet-E).}
  \label{fig:optimization}
\vspace{-0.4cm}
\end{figure}

In this section we aim to showcase the potential application of NDTs for optimization in QoS-aware scenarios. To this end, we use the \mbox{RouteNet-E} model used in the previous section. We define the optimization problem as finding the routing configuration that minimizes the average end-to-end delay on paths. We consider a destination-based OSPF routing scheme, where the initial routing configuration is the shortest path (i.e., equal weights on all links). To achieve optimization, we follow the reference workflow depicted in Fig.~\ref{fig:optimProcess}. Particularly, RouteNet-E represents the digital twin, while the network optimizer is implemented as an algorithm based on evolutionary strategies~\cite{wierstra2014natural}. In this architecture, the network optimizer generates variations of the shortest path (i.e., different link weights), and RouteNet-E is intended to predict the resulting delay on paths for those alternative configurations. Thus, the optimizer compares the delay predictions produced by RouteNet-E and it finally takes the routing configuration that results in minimum average end-to-end delay.

We evaluate the resulting optimizer in a synthetically generated topology of 25 nodes. 
Traffic matrices cover a wide range of traffic intensities (from low traffic load to highly congested networks). Figure~\ref{fig:optimization} shows the results of the optimization. Traffic intensity values (x-axis) represent the average traffic volume on paths (in bits per second). The final delay values (y-axis) are computed with the network simulator, used as ground truth in the previous section (see Table~\ref{tab:baslines}). As the network congestion increases (i.e., more traffic intensity), the network optimizer achieves higher delay reduction with respect to the initial shortest path. These results show that the NDT-based optimizer used in the experiments is able to effectively reduce the end-to-end delay on networks. 

\section{Open Challenges and Opportunities}

This article has discussed that modern ML techniques can be key enablers for building core components of NDTs, as well as described some potential applications of NDTs for a broad variety of networking use cases (see earlier). However, there are several open challenges that need to be addressed by the research community to enable the deployment of NDTs in real-world networks. Below we present some key open challenges and opportunities yet to be explored before achieving production-ready NDT solutions.

\textbf{\textit{Data collection and storage:}}
In a networking context, collecting and processing data is challenging and expensive. This is because it often requires the use of costly telemetry systems to gather relevant network state data. Moreover, data is only valuable if it has a common data format or labeling. However, in real-world networks data typically comes from different sources and has different formats. Thus, it is important to define standard representations and interfaces that can be applicable to different monitoring sources. 

One limitation of gathering network-related data is that it can require large amounts of storage. For example, in modern data center networks (with thousands of servers) most traffic flows have a very short duration~\cite{10.1145/1879141.1879175}. Therefore, to consider per-flow records can involve large amounts of data, making their storage and processing unfeasible.

In this context, monitoring platforms often use general-purpose compression methods to reduce the storage needs (e.g., GZIP). However, these methods do not exploit characteristics of traffic traces (e.g., temporal correlations), resulting in poor compression ratios. This calls for the design of special-purpose compression methods that can further help reduce the size of network monitoring data.

\textbf{\textit{Generalization and scalability to real networks:}}
The  NDT  should  be  able  to  perform well on different network scenarios than those seen during training. Generalization is important because training a NDT is not immediate, and network changes can happen very fast (e.g., link failure), so it is not possible to finish the training process before there is a new network event. One way to improve generalization of NDTs is using well-known ML techniques such as regularization or dropout. However, these methods can impact the performance or introduce a bias in the model. In addition, when the network scenario changes drastically, the NDT can lead to performance degradation. This presents an opportunity for the research community to develop new ML models that may lead to more solid generalization. In this context, GNN models have recently shown promising results for generalization across network-related data structured as graphs~\cite{ferriol2022routenet,eisen2020optimal}. 

Modern communication networks are often larger than the network environments used to generate the training datasets, raising a scalability challenge for ML models. NDTs should generalize well to networks considerably larger than those seen during training (e.g., 1-2 orders of magnitude larger). However, it often involves facing out-of-distribution values (e.g., larger traffic volumes and link capacities), which may degrade the performance of the NDT. Consequently, building scalable NDTs is an open issue that should be addressed to achieve production-ready solutions.

\textbf{\textit{Fine-grained control and management:}}
In order to perform efficient network operation it is necessary to model network traffic at a low granularity (e.g., flow-based operation). However, communication networks carry a large number of flows simultaneously~\cite{10.1145/1879141.1879175}, which may raise scalability issues for ML-based methods. Some networking systems tackle the flow scalability issue by applying traffic sampling or aggregation techniques. This enables the network operator to set a tradeoff between the sampling rate used and the accuracy of the statistics collected from the network. Therefore, building flow-based NDT models that can operate at a flow granularity and at short time scales is a relevant open challenge for the networking community.

\textbf{\textit{Dealing with uncertainty:}}
Neural network-based models are typically seen as a black-box, which hinders the deployment of DL solutions in real-world networks. When a neural network-based  model is evaluated, it is difficult to assess how certain the model is about the predictions made. Given the critical nature of communication infrastructures, such limitations are important as network operators need robust and reliable methods that can be applied to real-world networks without compromising their normal behavior. In this vein, existing works from the ML community attempt to solve this issue by modeling posterior probability distributions on DL models (e.g., Bayesian neural networks). Another alternative is to design comprehensive testing procedures in controlled network environments to systematically determine the safe operational ranges of DL models (e.g., supported traffic volumes), before deployment on customer networks.

\section{Conclusion}

This article has introduced the NDT concept and its reference architecture. We have argued that NDTs enable the development of more efficient network control and management tools for modern communication networks. In this context, recent advances in ML permit to build NDTs that can accurately mimic the behavior of real-world networks. In this article we focused on GNNs and DRL but we do not limit the application of other existing ML techniques to build market-ready NDTs. However, there are still some open challenges to be addressed for a full-scale NDT deployment in real networks. We encourage the networking community to explore innovative solutions to these challenges.

\vspace{-0.2cm}
\section*{Acknowledgment}

This publication is part of the Spanish I+D+i project TRAINER-A (ref.PID2020-118011GB-C21), funded by MCIN/ AEI/10.13039/501100011033. This work is also partially funded by the Catalan Institution for Research and Advanced Studies (ICREA) and the Secretariat for Universities and Research of the Ministry of Business and Knowledge of the Government of Catalonia and the European Social Fund.

\ifCLASSOPTIONcaptionsoff
  \newpage
\fi



%

\vspace{-0.4cm}

\bibliographystyle{IEEEtran}
\bibliography{references}

\vspace{-1.5cm}

\begin{IEEEbiographynophoto}{Paul Almasan}
is a PhD candidate at the Barcelona Neural Networking Center, Universitat Politècnica de Catalunya.
\end{IEEEbiographynophoto}
\vspace{-1.5cm}

\begin{IEEEbiographynophoto}{Miquel Ferriol-Galmés}
is a PhD candidate at the Barcelona Neural Networking Center, Universitat Politècnica de Catalunya.
\end{IEEEbiographynophoto}
\vspace{-1.5cm}

\begin{IEEEbiographynophoto}{Jordi Paillisse}
is a postdoctoral researcher at the Barcelona Neural Networking Center, Universitat Politècnica de Catalunya.
\end{IEEEbiographynophoto}
\vspace{-1.5cm}

\begin{IEEEbiographynophoto}{José Suárez-Varela}
is a postdoctoral researcher at the Barcelona Neural Networking Center, Universitat Politècnica de Catalunya.
\end{IEEEbiographynophoto}
\vspace{-1.5cm}

\begin{IEEEbiographynophoto}{Diego Perino}
is the Director of Telefónica Research, a team of researchers and technical experts in the areas of artificial intelligence, networks and systems, security and privacy and human computer interaction.
\end{IEEEbiographynophoto}
\vspace{-1.5cm}

\begin{IEEEbiographynophoto}{Diego López}
is a senior technology expert at Telefonica I+D, Chair of ETSI NFV and PDL ISGs.
\end{IEEEbiographynophoto}
\vspace{-1.5cm}

\begin{IEEEbiographynophoto}{Antonio Agustin Pastor Perales}
works as an expert for global technical areas in Telefónica I+D where he is involved in network and security innovation activities.
\end{IEEEbiographynophoto}
\vspace{-1.5cm}

\begin{IEEEbiographynophoto}{Paul Harvey}
is a research lead and Co-Founder of the Rakuten Mobile Innovation Studio.
\end{IEEEbiographynophoto}
\vspace{-1.5cm}

\begin{IEEEbiographynophoto}{Laurent Ciavaglia}
is a researcher at Rakuten Mobile and Co-Chair of the Network Management Research Group at the IETF.
\end{IEEEbiographynophoto}
\vspace{-1.5cm}

\begin{IEEEbiographynophoto}{Leon Wong}
is an industry research collaboration lead and research engineering lead for the Research and Innovation Lab in Rakuten Mobile.
\end{IEEEbiographynophoto}
\vspace{-1.5cm}

\begin{IEEEbiographynophoto}{Vishnu Ram}
is an independent researcher with more than two decades in the telecommunications industry.
\end{IEEEbiographynophoto}
\vspace{-1.5cm}

\begin{IEEEbiographynophoto}{Shihan Xiao}
received his PhD degree at Tsinghua University in 2017 and is currently a technical expert of Network AI at Huawei Technologies.
\end{IEEEbiographynophoto}
\vspace{-1.5cm}

\begin{IEEEbiographynophoto}{Xiang Shi}
is a senior engineer at the Network Technology Laboratory of Huawei Technologies.
\end{IEEEbiographynophoto}
\vspace{-1.5cm}

\begin{IEEEbiographynophoto}{Xiangle Cheng}
is a research fellow working on NetAI technologies at Huawei.
\end{IEEEbiographynophoto}
\vspace{-1.5cm}

\begin{IEEEbiographynophoto}{Albert Cabellos-Aparicio}
is a professor at Universitat Politècnica de Catalunya and director at the Barcelona Neural Networking Center.
\end{IEEEbiographynophoto}
\vspace{-1.5cm}

\begin{IEEEbiographynophoto}{Pere Barlet-Ros}
is a professor at Universitat Politècnica de Catalunya and scientific director at the Barcelona Neural Networking Center.
\end{IEEEbiographynophoto}
\vspace{-1cm}

\end{document}